# FEL INJECTOR CONTROL SYSTEM ON THE BASE OF EPICS

T.V. Salikova, A.A. Kondakov, G.Ya. Kurkin, A.D. Oreshkov, M.A. Scheglov, A.G. Tribendis
Budker Institute of Nuclear Physics, 630090, Novosibirsk, RUSSIA

## Abstract

The control system of the 1.5 MeV FEL injector is built on the base of ported EPICS. It uses low-cost hardware: personal computers with the processor Intel x86 and CAMAC equipment produced by our institute. At present time, the distributed control system includes one Pentium at OPerator Interface (OPI) level and two IOC (Input Output Controllers) under supervision of the real time operating system LynxOS/x86 at the low-level. Each IOC is used for monitoring of autonomous parts of the injector. The first IOC operates the Radio Frequency (RF) system. The second IOC operates the injector equipment.

## STATUS OF FEL

The first stage of the FEL (Free Electron Laser) complex [1] consists of a 1.5 MeV injector, one-track microtron-recuperator with accelerating RF system, and submillimeter FEL (Fig.1). The particle energy is 14 MeV. The bunch repetition rate is 0.022 to 22.5 MHz, and the average current is 10 to 50 mA. The FEL produces radiation of 1 to 10 kW average power and wavelength of 100 to 200 μm. The pulse duration is 20 to 100 psec. The first stage of the FEL is to test the recuperation of the beam power and to obtain the high power terahertz radiation.

At present time, the FEL injector and RF system were installed already, and installation of the microtron equipment is in progress.

The RF system contains one bunching cavity, two accelerating cavities fed by three generators of the injector, and 16 accelerating cavities fed by two generators of the microtron. The cavity has two contactless tuners of the fundamental-mode frequency, which is 180.4 MHz. The fundamental mode tuning range is 320 kHz. Two Higher-Order Modes (HOM) tuners detune the HOM frequencies while having almost no influence on the fundamental mode. The IOC of the RF system controls operation of all the cavities and RF generators. It also can run OPI functions for adjustment of the equipment. The RF system database includes 481 records.

The injector IOC operates the electrostatic gun, which emits electron bunches of 0.1 to 1.5 $n$sec duration with 0.022 to 22.5 MHz repetition rate. Power of a bunch reaches 300kV at the exit from the gun. Further, the electron bunch is accelerated in the RF cavities up to 1.5MeV. The injector IOC controls the injector magnetic system, which consists of solenoid lenses, correctors, pick-ups and beam transducers. The equipment for beam diagnostics is connected to the injector IOC. The injector system database includes upward of 150 records.

## FEL INJECTOR AUTOMATION

By now we have updated the software of the injector control system, which had been developed on the base of Windows95 with the LabWindows/CVI tool kit. Two variants of software for the RF control system have been developed: the first variant uses LabWindows/CVI, and the second variant is built on the base of EPICS. The RF system control program

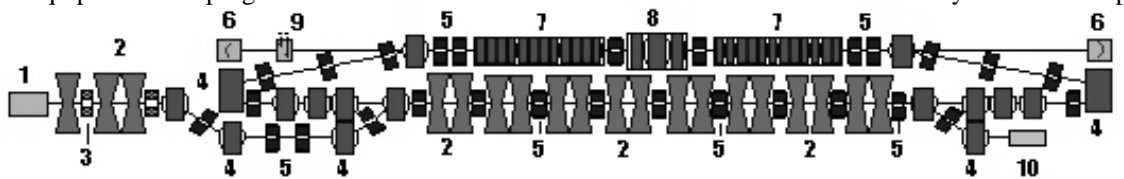

Fig.1. First stage of FEL.
1 – electron gun; 2 – RF cavities; 3 – solenoids; 4 – bending magnets; 5 – quadropole lenses; 6 – FEL optical resonator mirrors; 7 – undulators; 8 – buncher; 9 – outcoupler; 10 – beam dump under Windows95 executes a range of technological tasks, necessary during the commissioning and service of the equipment, and all functions of the control system. The EPICS variant of software executes only functions of the control system and is designed for the operation of the RF system as a part of the FEL complex.

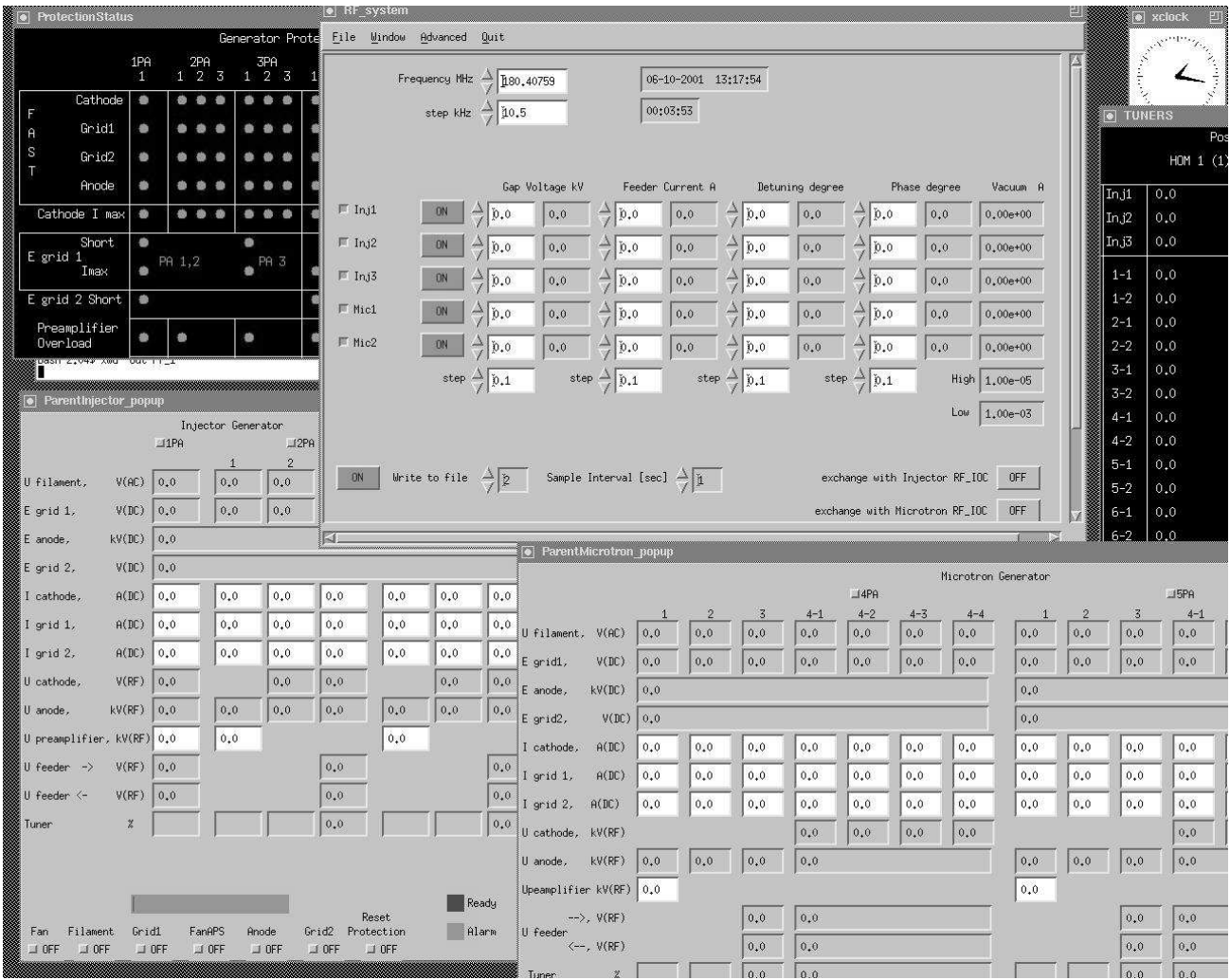

Fig.2 Main control panel of RF system.

The FEL control system is built on personal computers with Intel x86 processors under supervision of the real time operating system LynxOS, and CAMAC devices produced by our institute, which reduces the automation cost. The FEL control system uses 14 record types only and 20 device support routines, which were written for optimization of exchange with the equipment. All exchanges with the devices are executed by a standard mechanism, a driver supporting synchronous and asynchronous requests.

The CAMAC bus has low performance: the time of full execution of NAF reaches 16 μsec. Thus, control over the emergency conditions is executed by the hardware. That is the history-formed approach of our institute to solution of hard real time problems via creation of special hardware. A set of such devices was created for the FEL control system: devices for control over beam propagation in the injector, a gun driver, devices for control of locks of doors of the accelerator hall, a driver for monitoring of temperatures, etc.

At the OPI level, the RF control program (Fig.2) has seven panels for monitoring of the generator parameters of the injector RF system and microtron RF system as well as control panels for monitoring of three injector cavities and sixteen microtron cavities

The OPI program includes a mnemonic diagram of the injector, where each element is linked with its own control panel. A click to a mnemonic element generates a subwindow containing all control information.

The user has a set of windows made as a set of manipulation panels with mnemonic diagrams. A color palette indicates the alarm status. If a device fails, then the corresponding element in the mnemonic diagram is painted red. If a device works without errors, then its color is green.

The OPI programs of the RF system, gun and magnetic system log all updates of physical parameters to the cyclical buffer, which is periodically written to a file, which helps in analysis of failures of equipment.

## CONCLUSIONS

The FEL control system uses the low-cost hardware made by our institute for specific tasks of control systems of particle accelerators. EPICS guarantees a scalable system, which is important for development of the FEL control system. Performance of the EPICS tool kit meets the requirements of hard real time control.